# AN INTELLIGENT SYSTEM FOR CONTINUOUS BLOOD PRESSURE MONITORING ON REMOTE MULTI-PATIENTS IN REAL TIME

Roberto Marani and Anna Gina Perri

Electrical and Information Engineering Department, Electronic Devices Laboratory,
Polytechnic University of Bari, via E. Orabona 4, Bari - Italy
perri@poliba.it

*ABSTRACT*

*In this paper we present an electronic system to perform a non-invasive measurement of the blood pressure based on the oscillometric method, which does not suffer from the limitations of the well-known auscultatory one. Moreover the proposed system is able to evaluate both the systolic and diastolic blood pressure values and makes use of a microcontroller and a Sallen-Key active filter. With reference to other similar devices, a great improvement of our measurement system is achieved since it performs the transmission of the systolic and diastolic pressure values to a remote computer. This aspect is very important when the simultaneous monitoring of multi-patients is required. The proposed system, prototyped and tested at the Electron Devices Laboratory (Electrical and Information Engineering Department) of Polytechnic University of Bari, Italy, is characterized by originality, by plainness of use and by a very high level of automation (so called "intelligent" system).*

## 1. INTRODUCTION

The most recent developments of electronics, informatics and telecommunications let consider applications in the biomedical engineering field to improve the healthcare quality [1-8]. In particular a number of systems has been developed in the telemedicine and home care sectors which could guarantee an efficient and reliable home assistance allowing a highly better quality of life in terms of prophylaxis, treatment and reduction of discomfort connected to periodic out–patient controls and/or hospitalization for the patients afflicted by pathologies, such as hypertension, and allowing considerable savings on sanitary expenses.

In particular hypertension is defined as elevated blood pressure (BP) above 140 mm Hg systolic and 90 mm Hg diastolic when measured under standardized conditions [9] [10]. Hypertension can be a separate chronic medical condition estimated to affect a quarter of the world's adult population [11], as well as a risk factor for other chronic and non-chronic patients. Traditional high-risk patients include all patients afflicted by pathologies such as cardiac decompensation, ischemic heart desease, kidney disease, diabetes. Persistent hypertension is one of the key risk factors for strokes, heart attacks and increased mortality [12]. In particular in pregnant women with gestational diabete, known as preeclampsia, hypertension is the most common cause of maternal and fetal death [13]. For all the previous cases blood pressure should be kept below 130 mmHg systolic and 80 mm Hg diastolic to protect the kidneys from BP-induced damage [14].

Therefore, in particular for high-risk patients, it is very important to employ a system for monitoring blood pressure over a long period (for example, of twenty-four hours), without compromising the ordinary day activities.

The most used method for measuring blood pressure is the auscultatory method, which involves an operator besides the same patient. This method is based on the contemporary use of a sphygmomanometer and a stethoscope. The sphygmomanometer has a cuff, which inflates and deflates, equipped with a pressure sensor positioned on the arm in correspondence of the brachial artery. The stethoscope allows to listen to arterial sounds (known as Korotkoff sounds) during the cuff slow deflaction, which are used to determine systolic and diastolic blood pressure. This method presents some difficulties in signal analysis due to physiological variations of the Korotkoff sound patterns; moreover, weak signals are disturbed by ambient noises and a misleading information can occur [15]. Moreover it is not possible to have the pressure data in real time and the simultaneous monitoring of a group of patients.

In place of the auscultatory method, there are other important indirect methods such as the oscillometric method, which is one of the best approach to evaluate the systolic and diastolic blood pressure [16].

In this paper we propose an electronic system to perform a non-invasive measurement of the blood pressure based on the oscillometric method and able to evaluate both the systolic and diastolic blood pressure values. With reference to other similar devices, a great improvement of our measurement system has been achieved since it performs the transmission of the systolic and diastolic pressure values to a remote computer. This aspect is very important when the simultaneous monitoring of multi-patients is required.

The proposed system, prototyped and tested at the Electron Devices Laboratory (Electrical and Information Engineering Department) of Polytechnic University of Bari, Italy, is characterized by originality, by plainness of use and by a very high level of automation (so called "intelligent" system). The paper is organized in the following way: in Section 2 we describe the main features of the proposed system respect to the state-of-art, while in Section 3 the proposed sensor architecture is analyzed. In Section 4 we describe the transmission architecture, highlighting the main goals obtained by our design. Finally the conclusions and future developments have been illustrated in Section 5.

## 2. MAIN FEATURES OF THE PROPOSED SYSTEM

The actual systems are inadequate for continuous blood pressure monitoring, not easy to use and have to be managed only by qualified operators, which makes them unsuited for personal use and domestic applications. Moreover one of the limitations of existing devices lies in the fact that they are not wearable.

Our measurement system overcomes the previous limitations and it has been designed to be employed also to real-time rescue in case of emergency without the necessity for data to be constantly monitored by a medical centre, leaving patients free to move. This system is particularly devoted to high-risk patients i.e. all patients afflicted by pathologies such as cardiac decompensation, ischemic heart desease, kidney disease, diabetes, improving patient's safety and quality of life.

The proposed system is based on the oscillometric method [16]. This approach analyzes the variations in pulse pressure as a function of the pressure applied to a pneumatic cuff wrapped around the limb. As in the auscultatory method, the cuff is inflated until the artery is completely occluded. A stepwise decrease in cuff pressure is then applied, and an increase in pulse amplitude is observed when the cuff pressure equals the blood systolic pressure. The pulse amplitude increases until the mean blood pressure is reached. The pulse amplitude then decreases with decreasing of the cuff pressure from mean to diastolic values. The systolic and diastolic blood pressure is then evaluated by applying a suitable numerical algorithm to the shape of oscillometric amplitudes.

Moreover, this method allows the measurement of the blood pressure also when the Korotkoff sounds are weak, thus overcoming the limitations related to the auscultatory method. An important aspect is that no microphone is required for sensing.

The architecture of our device is shown in Fig.1.

It consists of the following main sections:

block 1.1 is the pressure transducer;

block 1.2 is the signal conditioning circuit, which includes a band-pass filter for the extrapolation of the oscillations and an analog to digital converter (ADC) for signal acquisition output to the transducer and its filtered version;

block 1.3 is the intelligent and programmable section, based on a microcontroller for the signal processing: after deflation of the cuff; it allows to determine the peak of the pulsatile component and the diastolic and systolic pressure as a percentage of that maximum;

block 1.4 is the memory unit;
block 1.5 is the wireless transmitter and receiver unit;
block 1.6 is the power supply circuit.

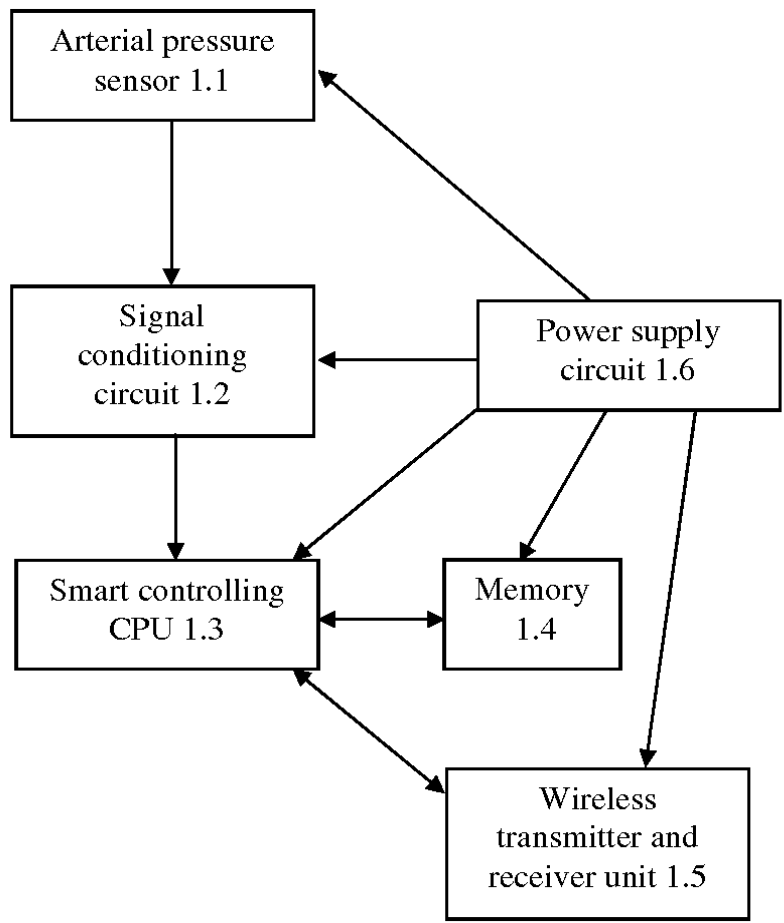

Figure 1. Bloch scheme of the proposed device.

Moreover our systems has been designed to be employed also for real-time rescue in case of emergency without the necessity for data to be constantly monitored by a medical centre, leaving patients free to move. This possibility is particularly devoted to high-risk patients, as we have already explained. In fact, the combination of the latest suitable telecommunication solutions (GPRS and Bluetooth) with new algorithms and solutions for automatic real-time diagnosis, cost-effectiveness (both in terms of purchase expenses and data transmission/analysis) and simplicity of use (the patient will be able to wear it) can give the designed system useful for remote monitoring, allowing real-time rescue operations in case of emergency without the necessity for data to be constantly monitored.

For this purpose the proposed system has been equipped with properly developed firmware [1-2] [17], which enables automated functioning and complex decision-making. It is indeed able to prevent lethal risks thanks to an automatic warning system. All this occurs automatically without any intervention of the user.

Each monitored patient is given a case sheet on a Personal Computer (PC) functioning as a server (online doctor). Data can also be downloaded by any other PC, palmtop or smart phone equipped with a browser. The system reliability lies on the use of a distributed server environment, which allows its functions not to depend on a single PC and gives more online doctors the chance to use them simultaneously.

The system consists of three hardware units and a management software properly developed. The units are:

- the sensors for the measurement of blood pressure;
- a Portable Unit (PU), which is wearable and wireless (GPRS/Bluetooth). This PU allows, by an Internet connection, the transmission, continuous or sampled or on demand, of the health parameters and allows the GPS satellite localization and the automatic alarm service. Moreover PU has an USB port for data transfer and a rechargeable battery;
- Relocable Unit (RU): GPRS/Bluetooth Dongle (on PC server, i.e. online doctor).
- Management Software: GPS mapping, address and telephone number of nearest hospital, simultaneous monitoring of more than one patient, remote (computerized) medical visits and consultation service, creation and direct access to electronic case sheets (login and password)

Fig. 2 shows a picture of the PU. The very small dimensions are remarkable, even if it is only a prototype, realized at the Electron Devices Laboratory of Polytechnic University of Bari, and more reduction in dimensions is still possible.

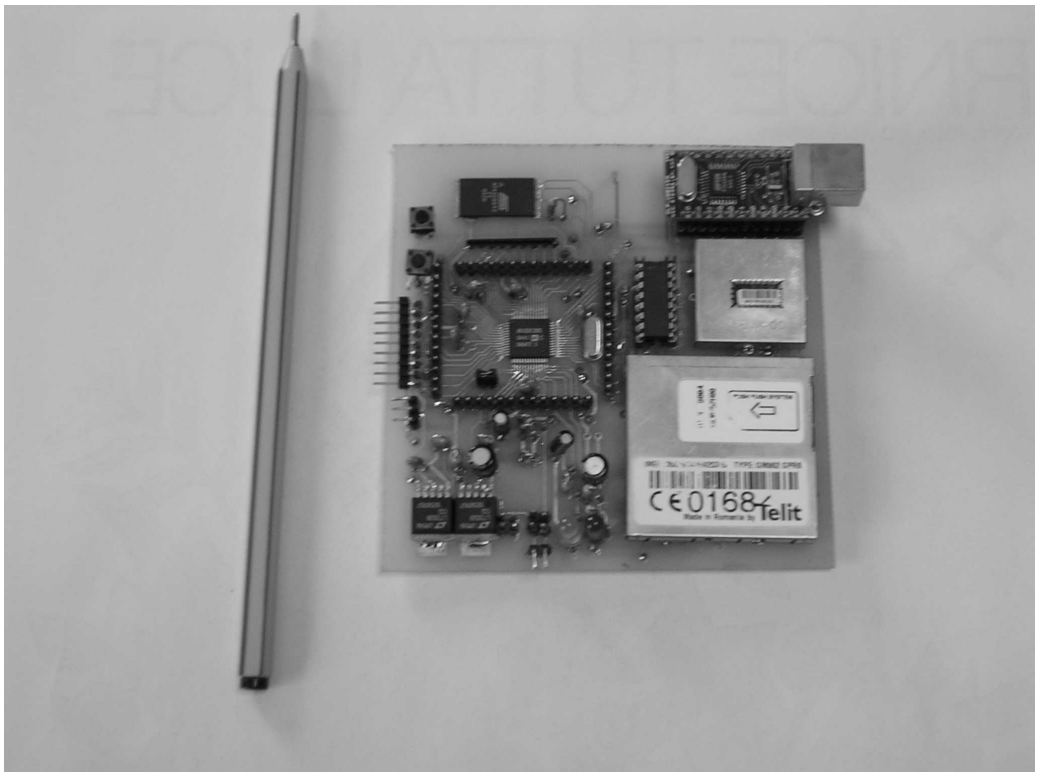

Figure 2. The prototype of the Portable Unit.

The system, in particular the PU, collects data continuously. These are stored in an on-board flash memory and then analyzed in real-time by an on-board automatic diagnosis software. Data can be sent to the local receiver, directly to the PC server (online doctor), or to an Internet server, which allows anyone to download them once identified with his/her own login and password.

Data can be transmitted as follows:
1. real time continuously
2. at programmable intervals (for 30 seconds every hour, for example)
3. automatically, when a danger is identified by the alarm system
4. on demand, whenever required by the monitoring centre
5. offline (not real-time), downloading previously recorded (over 24 hours, for example) data to a PC.

In all cases patients do not need to do anything but simply switching on.

When an emergency sign is detected through the real time diagnosing system, the PU automatically sends a warning message, indicating also the diagnosis, to one person (or even more) who is able to verify the patient health status and arrange for his/her rescue. In order to make rescue operations as soon as possible, the PU provides the patient's coordinates using the GPS unit and the Management Software provides in real time a map indicating the position of the patient, as shown in Fig. 3.

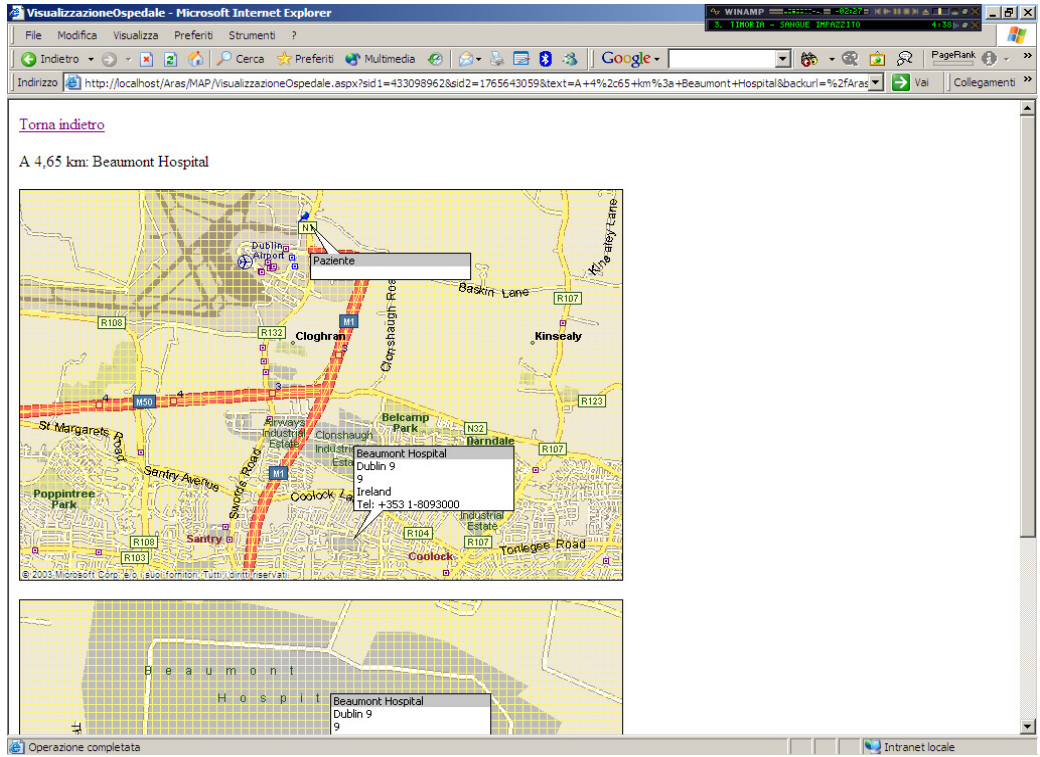

Figure 3. GPS mapping with address and telephone number of nearest hospital.

## 3. ANALYSIS OF THE PROPOSED SENSOR ARCHITECTURE

With referring to Fig. 1, the arterial pressure, coming from an inflated cuff, is firstly transduced into a voltage signal and, after a pass band filtering stage, is processed by a microcontroller which applies a numerical algorithm to achieve the systolic and diastolic values of blood pressure that are then transmitted.

A transducer is required to transform the pressure into a voltage signal. We have chosen a piezoresistive transducer for its high sensitivity. These transducers are based on the piezoresistive effect, which is the change of the electric resistance when a mechanical stress occurs. The change in resistance is described by the following relationship:

$$\frac{\Delta R}{R_0} = \frac{\Delta \rho}{\rho_0} + (1 + 2\mu)\frac{\Delta L}{L_0} \qquad (1)$$

where $\mu$ is the Poisson coefficient, while $R_0$, $\rho_0$, $L_0$ are the electric resistance, the resistivity and the length of the unloaded sensor, respectively. $\Delta L$ is due to the breast dilatation.

Most of piezoresistive transducers used in medical applications are based on semiconductor materials, in which $\Delta\rho/\rho_0$ is greater than other terms in (1). Moreover, these active sensors are arranged as a Wheatstone bridge, and provide an output voltage given by:

$$\Delta V = \frac{\Delta R}{R_0} V_S \qquad (2)$$

where $V_S$ is the supply. The output signal is of about mV.

We have chosen a piezoresistive transducer constituted by two gain stage, the first of them arranged to provide also a temperature compensation (Fig. 4).

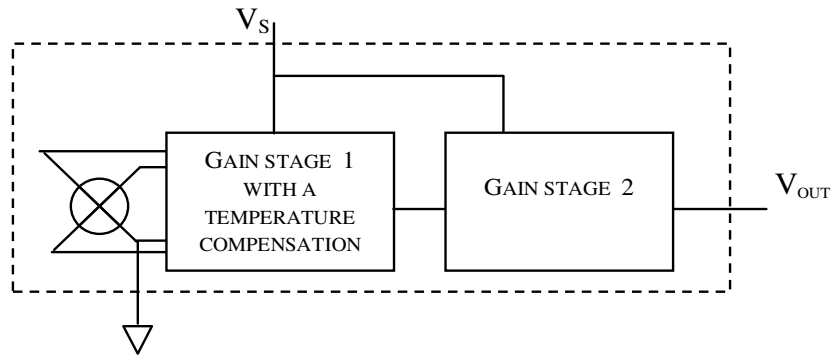

Figure 4. Two-stage pressure sensor.

The sensor is characterized by an input pressure range of $0 \div 50$ kPa, an output voltage range of $0.2 \div 4.7$ V, a sensitivity of 90 mV/kPa = 12 mV/mmHg, a supply voltage of 5V and supply current of 7 mA. The signal coming from the piezoresistive transducer is the superposition of a deflation signal with the arterial pressure pulsation one, the former being a low frequency signal. The systolic and diastolic blood pressure values can be evaluated by using a numerical algorithm based on a comparison between the oscillometric and the deflation signal. To separate the deflation signal from the oscillometric one a passband digital filter with a low cut-frequency is required.
To properly design the filter, we have firstly analyzed the composed signal acquired at 100 samples per second in a 100 seconds measurement process, thus obtaining 10000 samples.
In Fig. 5 the signal under investigation is shown. In this figure two curves can be distinguished, corresponding to inflating and deflating processes, respectively. We analyze the signal during the deflating process, which occurs after the first six seconds. The application of the Fast Fourier Transform algorithm allows to analyze the frequency components of the signal, as depicted in Fig. 6.

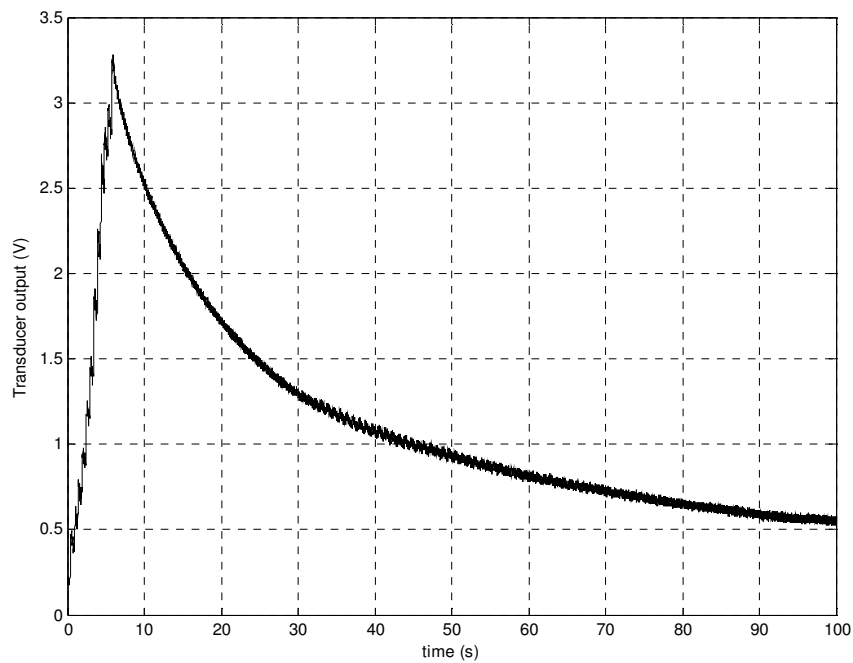

Figure 5. Pressure signal behaviour.

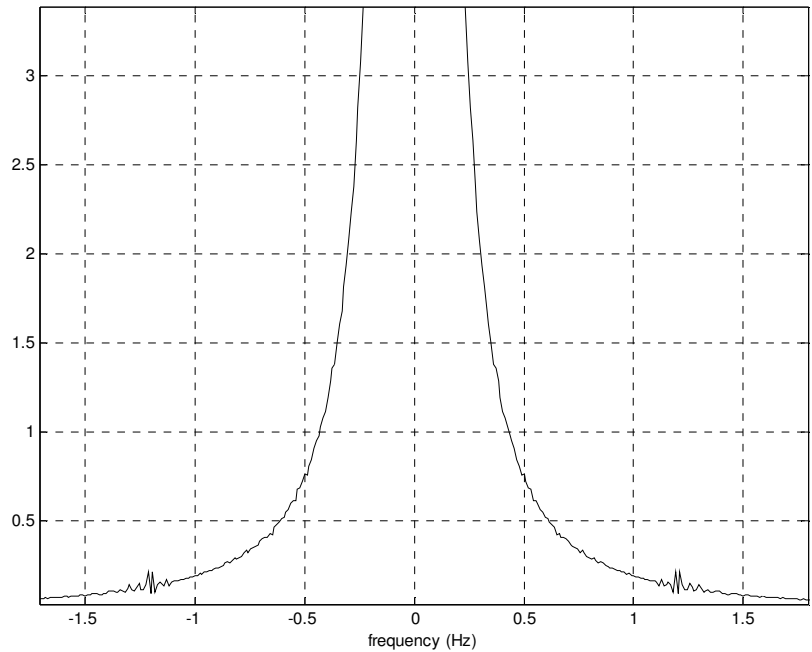

Figure 6. Pressure signal spectral components.

The zero frequency component, which is due to non-zero mean value of the deflating pressure signal, is predominant with respect to the other frequency components located in the 1 ÷ 1.5 Hz range and due to the pulsation of the blood flow corresponding to the heart beat period. In order to eliminate the zero component frequency, the 50 Hz noise and the thermal noise at high frequencies, a pass-band filter is required. We have designed a second-order narrow-band Sallen-Key active filter, constituted by cascade of a low-pass filters and high-pass, and a gain stage. The low-pass filter architecture is represented in Fig. 7, where R = 33 kΩ, C = 100 nF, thus giving a central frequency $\omega_0$ = 1/RC = 303 rad/s and a cut-off frequency $\omega_H$ = 195 rad/s (i.e. $f_H$ = 31 Hz). The high-pass filter can be easily obtained by exchanging the resistor and capacitor, choosing R = 56 kΩ and C = 2.2 nF thus giving $\omega_0$ = 8.1 rad/s and $\omega_L$ = 12.6 rad/s (i.e. $f_L$ = 1.29 Hz).

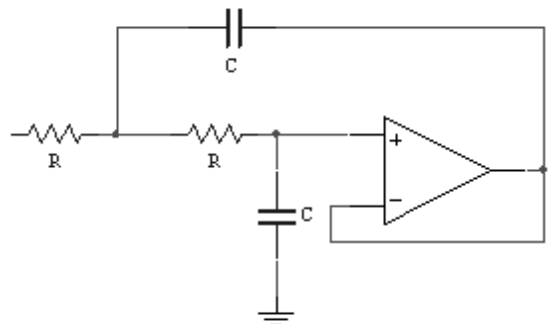

Figure 7. Low-pass active filter.

Since the medium arterial pressure is typically 2 mmHg, we obtain, with a sensor having a sensitivity of 12 mV/mmHg, an output voltage signal of 24 mV.
The last amplifier has been designed to provide a gain of 172, which is suitable to give an output voltage of 4.5 V, compatible with the technical requirement of ADuC812 microcontroller. The filter is shown in Fig. 8, while the voltage reference of 1.16 V is realized as in Fig. 9.

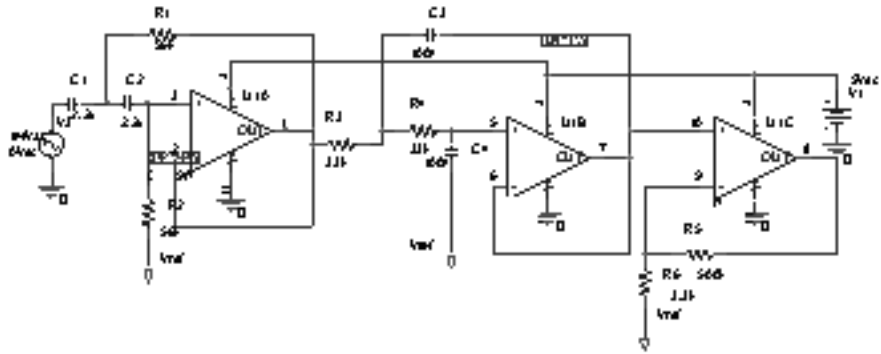

Figure 8. Pass-band Sallen-Key active filter.

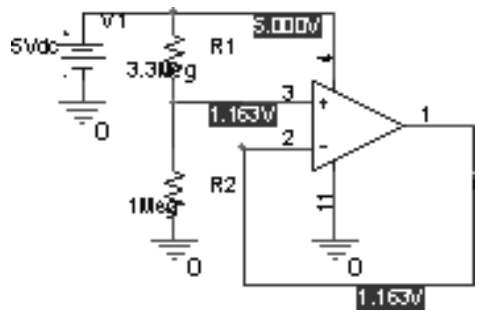

Figure 9. Voltage reference.

The frequency response of the designed pass-band filter is shown in Fig. 10.

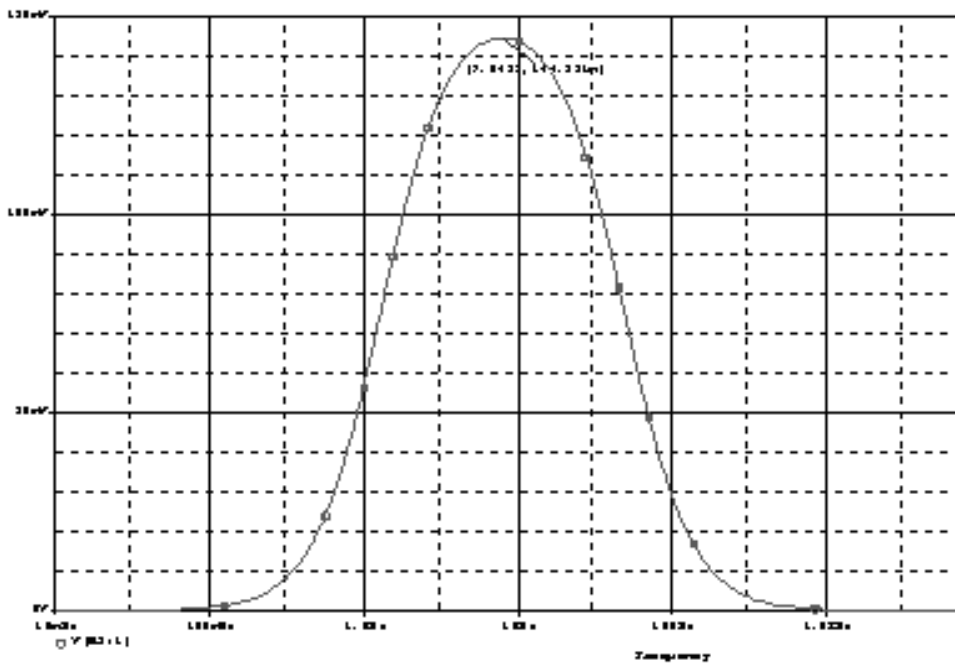

Figure 10. Frequency response of the active filter.

Fig. 11 shows the oscillometric signal, which is obtained by applying the narrow pass-band active filter to the pressure signal coming from the sensor-transducer system.

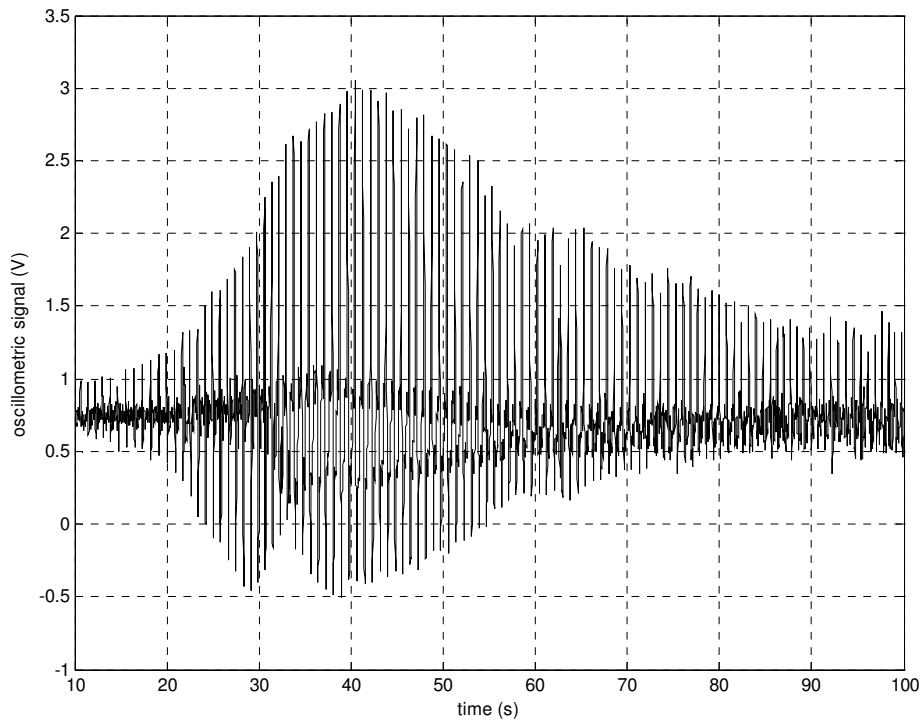

Figure 11. Oscillometric signal.

## 4. ANALYSIS AND TEST OF THE TRANSMISSION ARCHITECTURE

In order to evaluate the systolic and diastolic pressure values, a numerical algorithm has to be applied. This task is carried out by a microcontroller which compares the signal from the transducer with the output signal of the pass-band filter, thus evaluating the mean arterial pressure and the related values corresponding to systolic and diastolic blood pressure. We have used an Analog Devices AduC812 microcontroller, characterized by analog single-ended 8 acquisition channels, each of them equipped by a 12 bit ADC converter, flash and central memory, an UART serial interface. We use only two channel to process the input signals, which are sampled at different frequencies, stored in the internal memory and then compared. We have written an assembly code to control each function of the AduC microcontroller.

The main aspects of our design is the possibility to transmit the recovered data to a remote computer through a wireless communication system. To this purpose we have used a transmitting module employing a surface acoustic wave (SAW) transmitter to produce a carrier wave at the free frequency of 433.92 MHz at a voltage supply of 5 V and an absorbed current of 4 mA.

The chosen receiver is characterized by a high selectivity and insensitivity to electromagnetic fields, a working centre frequency of 433.92 MHz and a pass-band of 600 kHz. The typical values of the voltage supply and absorbed current are +5 V and 3 mA respectively. Moreover, the receiver is characterized by a high sensitivity of -100 dBm which can be lowered in order to provide a suitable noise level reduction. The noise level reduction can be reached by connecting a suitable resistance between a proper pin and the ground. We have chosen a resistance R = 1 MΩ which exhibits a gain loss of 1dB. The receiver can operate only with pass-band signals, and the minimum frequency is 100 Hz. Therefore, a signal having an amplitude constant for a duration greater than 0.01 s could not be correctly recognized.

To solve this problem we have arranged the transmission circuit as shown in Fig. 12, in which a p-n-p transistor having a low $V_{CEsat}$ voltage (- 90 mV) and a base-emitter voltage $V_{BE} = 0.7V$ has been used.

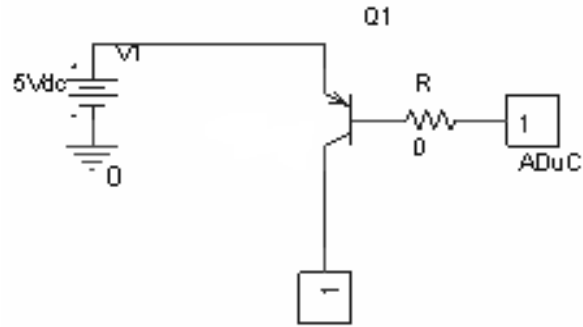

Figure 12. Transmission circuit arrangement.

When the output pin of the ADUC512 is held at low level, the BJT is turned on and the transmitter is powered on.
In order to allow a synchronization between the transmitter and the receiver, a short pulse is sent to the transmitter just before starting the transmission of the modulated data. This pulse is obtained activating the transmitter for a very short time, less than 0.01 s, using the circuit of Fig. 12.
Since the transmitter exhibits an input resistance depending on the state (ON or OFF) of the device, we have then uncoupled the two stage, microcontroller and transmitter, using an OPAMP-based buffer stage, shown in Fig. 13.

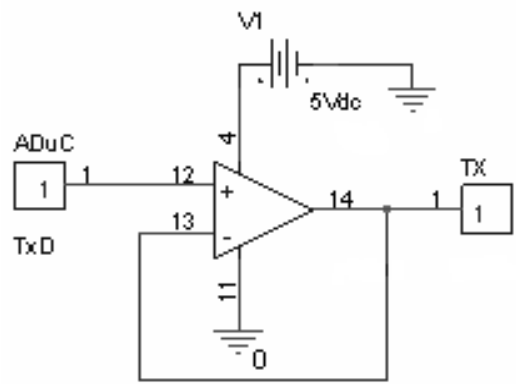

Figure 13. Buffer stage between AduC812 and transmitter.

The complete circuit is reported in Fig. 14.

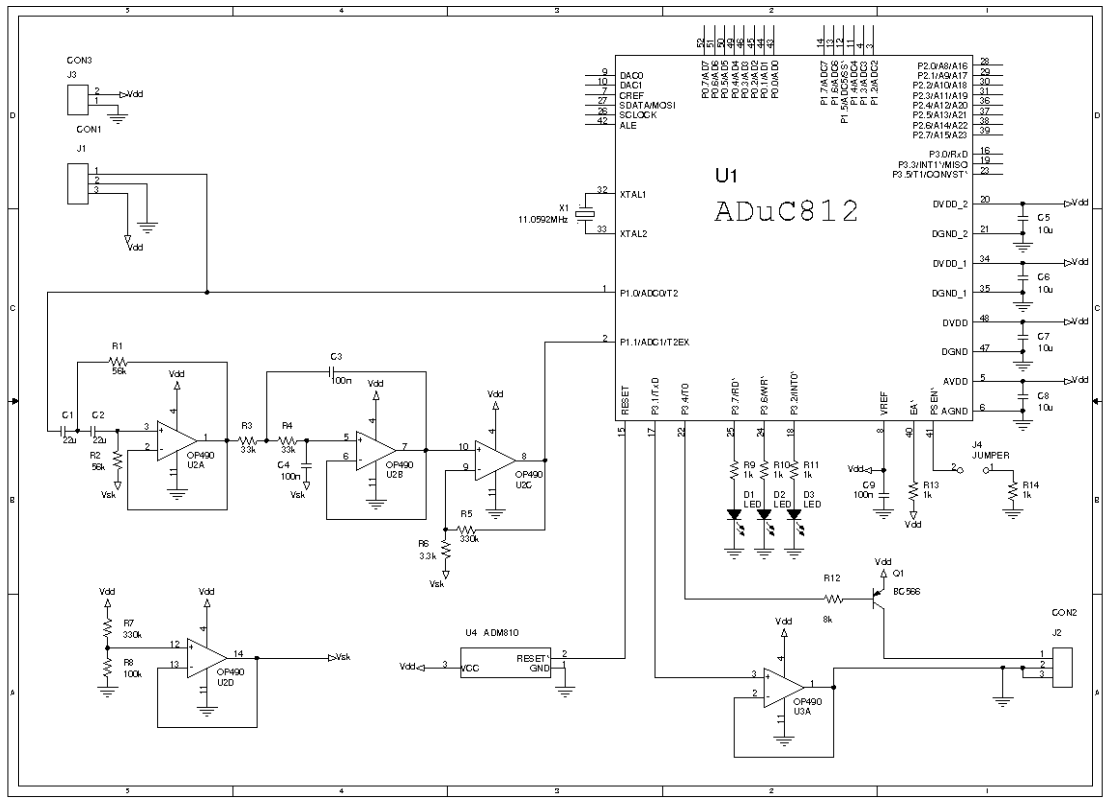

Figure 14. The complete scheme of the proposed system.

Finally, we have tested our system by using a Lu-La Logic Analyzer, which has two input channel: we have connected the channel 0 to the receiver output and the channel 1 to the transmitter input, as shown in Fig. 15.

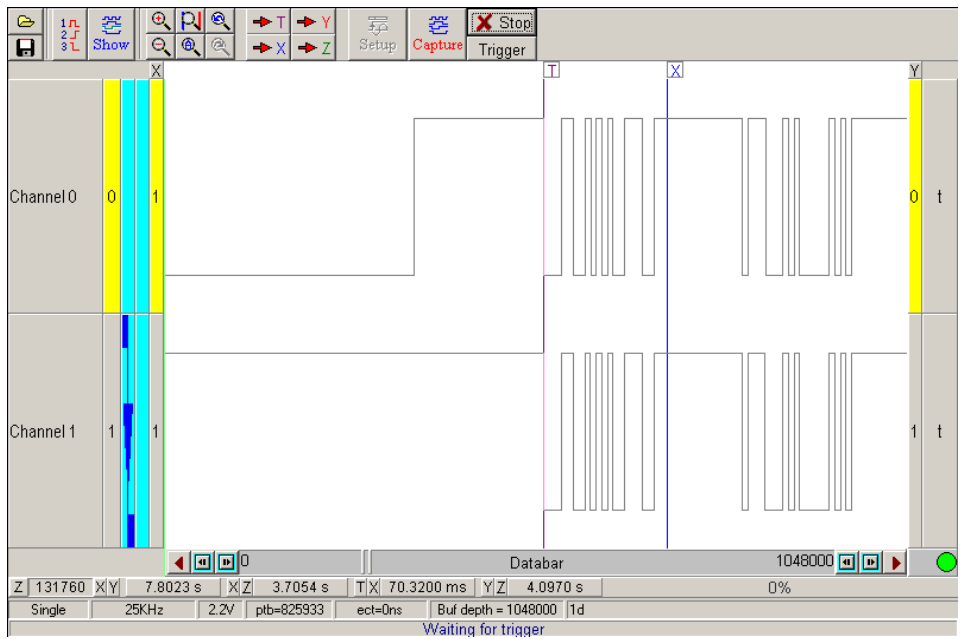

Figure 15. Testing stage by using a Lu-La Logic Analyzer. Channel 0 and 1 show the receiver output and transmitter input, respectively .

As shown in Fig. 15, the transmitted signal is perfectly recognized by the receiver.

## 5. CONCLUSIONS AND FUTURE DEVELOPMENTS

In this paper a system to measure the arterial blood pressure has been presented. Our device is based on the use of the oscillometric method, which solves the all the problems related to the typical approach detecting the Korotkoff sounds. The presented architecture makes use of a second order narrow band active filter and a microcontroller ADuC812, which applies a numerical approach to evaluate the systolic and diastolic values of the arterial pressure. Moreover, our device can perform the transmission of the measured pressure values to a remote computer using a working centre frequency at 433.92 MHz. This feature, which is very promising for the simultaneous monitoring of several patients, will be further improved in the future developments by using a GPRS network.

Authors

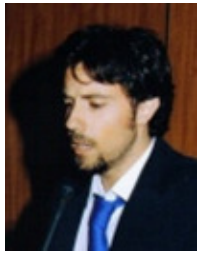

**Roberto Marani** received the Master of Science degree (cum laude) in Electronic Engineering in 2008 from Polytechnic University of Bari, where he received his Ph.D. degree in Electronic Engineering in 2012.

He worked in the Electronic Device Laboratory of Bari Polytechnic for the design, realization and testing of nanometrical electronic systems, quantum devices and FET on carbon nanotube. Moreover Dr. Marani worked in the field of design, modelling and experimental characterization of devices and systems for biomedical applications.

In December 2008 he received a research grant by Polytechnic University of Bari for his research activity. From February 2011 to October 2011 he went to Madrid, Spain, joining the Nanophotonics Group at Universidad Autónoma de Madrid, under the supervision of Prof. García-Vidal.

Currently he is involved in the development of novel numerical models to study the physical effects that occur in the interaction of electromagnetic waves with periodic nanostructures, both metal and dielectric. His research activities also include biosensing and photovoltaic applications.

Dr. Marani is a member of the COST Action MP0702 - Towards Functional Sub-Wavelength Photonic Structures, and is a member of the Consortium of University CNIT – Consorzio Nazionale Interuniversitario per le Telecomunicazioni.

Dr. Marani has published over 100 scientific papers.

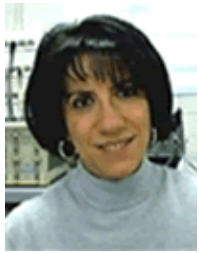

**Anna Gina Perri** received the Laurea degree cum laude in Electrical Engineering from the University of Bari in 1977. In the same year she joined the Electrical and Electronic Department, Polytechnic University of Bari, where she is Professor of Electronics from 2002.

Her current research activities are in the area of numerical modelling and performance simulation techniques of electronic devices for the design of GaAs Integrated Circuits and in the characterization and design of optoelectronic devices on PBG. Moreover she works in the design, realization and testing of nanometrical electronic systems, quantum devices, FET on carbon nanotube and in the field of experimental characterization of electronic systems for biomedical applications..

Prof. Perri is the Head of Electron Devices Laboratory of the Polytechnic University of Bari. She is author of over 250 book chapters, journal articles and conference papers and serves as referee for many international journals.

Prof. Perri is a member of the Italian Circuits, Components and Electronic Technologies – Microelectronics Association and is a member of the Consortium of University CNIT – Consorzio Nazionale Interuniversitario per le Telecomunicazioni.

Prof. Perri is a Member of Advisory Editorial Board of International Journal of Advances in Engineering & Technology and of Current Nanoscience (Bentham Science Publishers).